\documentclass[aps,twocolumn,showpacs,english]{revtex4-1}

\usepackage[T1]{fontenc}
\usepackage[latin9]{inputenc}
\usepackage{textcomp}
\usepackage{amsmath}
\usepackage{graphicx}
\usepackage{amssymb}
\usepackage{babel}
\begin{document}

\title{Probing photon energy statistics of an emitter using photon correlations}

\author{A. Trichet$^{1*}$, S. Bounouar$^{1,2*}$}

\affiliation{
$^1$ CEA-CNRS-UJF group 'Nanophysique et Semiconducteurs', Institut N\'{e}el, CNRS - Universit\'{e} Joseph Fourier, 38042 Grenoble, France, \\
$^2$ CEA-CNRS-UJF group 'Nanophysique et Semiconducteurs', CEA/INAC/SP2M, 38054 Grenoble, France}

\begin{abstract}
We investigate the possibility of measuring the homogeneous and inhomogeneous contribution to the linewidth of a spectrally diffusing single photon emitter using a simple photon correlation spectroscopy method (PCS). The photon energy statistics of the homogeneous line (poissonian) and of the spectral diffusion (first order markovian) being of different natures, they act differently on the half-line autocorrelation function (HLAF). We model here their effects and show it is possible to extricate them, opening the opportunity to determine separately the homogeneous linewidth, and the spectral diffusion amplitude.
\end{abstract}


\maketitle

\section{Introduction}

It is known since the pioneer works of Handbury-Brown and Twiss on the correlation of an optical field \cite{hbt}, and of Kimble, Dangennais and Mandel on the correlation of a quantum state of the light \cite{ kimb1}, that it is possible to probe the emission statistics of an emitter by photon correlations. In particular Kimble et. al showed that for a two levels system, the autocorrelation function of the  emitted optical field increases with the delay $\tau$  \cite{ kimb2}. This is a characteristic of a single photon emission. 
 In condensed matter systems, the energies of the emitted single photons can fluctuate.  The interaction of the emitter with the numerous degrees of freedom of the surrounding environment leads to random fluctuations of the emission energy. Since quantum computation requires indistinguishable photons  \cite{ knill} in order to proceed to efficient two-photons interferences \cite{hongou}, this is a severe limitation in the use of solid state systems for such operations. 
This dephasing phenomen is the result of several different processes. From the energy statistics point of view, one can divide them in two categories: Poissonian processes, such as exciton-phonon coupling, randomly affect the energy of the emitted photons. They are "memoryless" processes as the energy of the emitted photon at time t does not depend on the energy of the emitted photon at time $t-\tau$ whatever the value of $\tau$. We define the energy distribution of these uncorrelated random processes as the homogeneous linewidth of the emitter. On the other hand, a correlated process can shift the emitter energy. Fluctuations of the electronic environment are also a source of dephasing. This random Stark-shift of the emission energy is called spectral diffusion \cite{empedocles}. In opposition with phonon coupling, spectral diffusion is
time-correlated and it is is a reasonable approximation to consider it as a first order Markovian process with a correlation time $\tau_c$ \cite{markovian, markovian2}. Unlike the poissonian process, which is independent of its history, the first order markovian
process is influenced by its immediate or most recent past. Thus, the energy of the emitted photon shifted by the spectral diffusion depends on the  energy of the previously emitted photon.
 In a previous publication \cite{gregdiff}, it was shown that, by taking the Handbury-Brown and Twiss setup and by adding detection energy conditions, such as selecting only one half of the emission line, one has access to the subnanosecond correlation time of the spectral diffusion $\tau_c$ \cite{gregdiff}. In this letter, we show that this measurement technique is even more powerful since the resulting half line autocorrelation function (HLAF) brings informations on the relative contribution of the correlated and uncorrelated processes. Thus, such a slight modification of the original HBT setup, which was providing emission statistics of the emitter, gives as well access to the energy statistics of the emitter. 
In a first part, we describe the main principles of the experiment. Then, 
after showing that it is always possible to separate the emission statistics and the energy statistics contributions on the HLAF, we derive analytically the HLAF of a spectrally diffusing single photon emitter with a finite homogeneous linewidth. After discussing its properties, we confirm the validity of our model with a Monte Carlo simulation, building numerically the HLAF of a spectrally diffusing poissonian emitter and simulating the influence of a finite homogeneous linewidth. To finish, we highlight  the possible
 direct application of this theoretical result: the separate determinations of the homogeneous linewidth and the spectral diffusion amplitude.

\section{Principle of the measurement}

\begin{figure}[h]
\noindent \centering{}\includegraphics[scale=0.4]{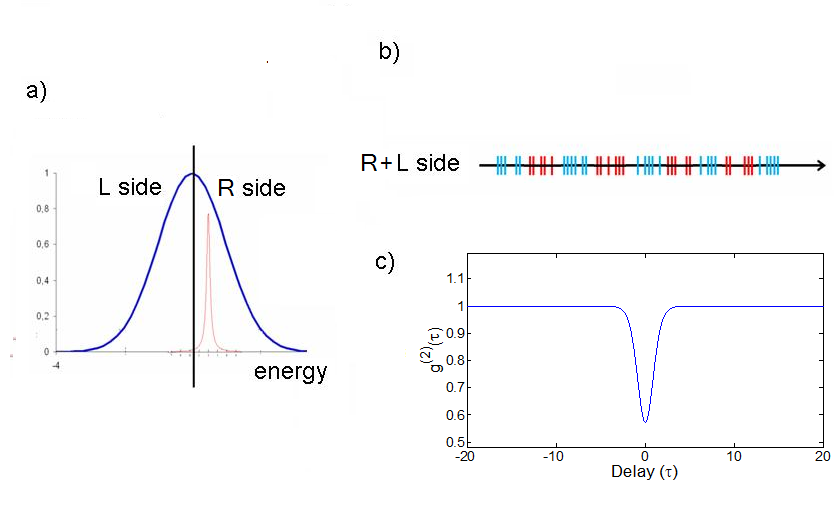}\caption{a) spectrum of a spectrally diffusing emitter. b) photons statistic of the emitter in the HBT experiment c) Autocorrelation function of the single photon emitter convoluted with the temporal resolution.}\label{singlephoton}
\end{figure}

The principle of the PCS experiment is to convert fluctuations of the emitter energy in intensity fluctuations. In fig. \ref{singlephoton} a) we consider the situation where all the photons are collected and sent to the HBT setup whatever their energies (photons comming from the right side of the line are coloured in red, from the left side in blue). The scheme in fig. \ref{singlephoton} b)  represents the single photon emission with photons arriving one by one with a characteristic time defined by the emitter radiative lifetime. The HBT setup allows to measure the intensity first order correlation function \cite {mandel}: 

\begin{equation}
g^{2}(\tau)=\frac{\langle I(0)I^*(\tau)\rangle}{\langle I(\tau)\rangle^2}
\end{equation}

This function is plotted in fig. \ref{singlephoton} c). The dip at zero delay is the signature of single photon emission. It is non-null at zero delay only because of the convolution with time resolution of the photodetectors.

\begin{figure}[h]
\noindent \centering{}\includegraphics[scale=0.4]{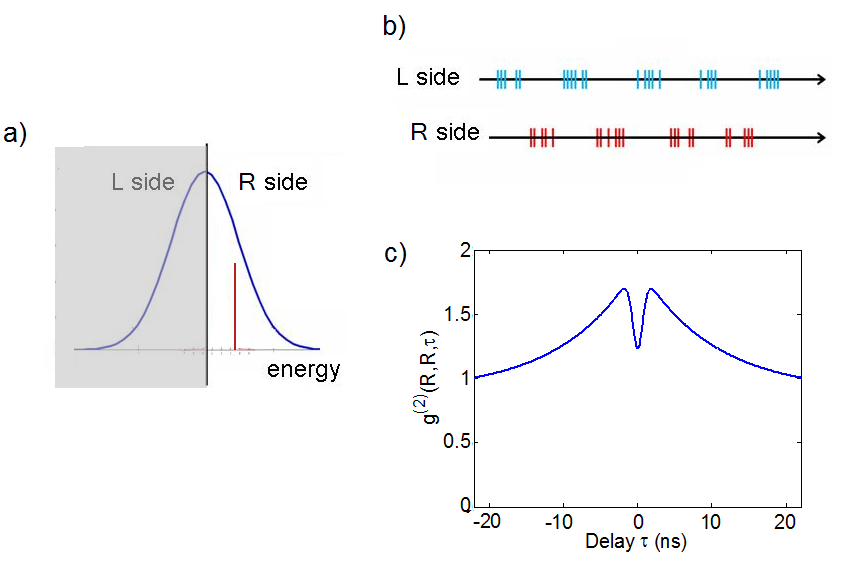}\caption{a) spectrum of a spectrally diffusing emitter with an infinitely small linewidth: only the right spectral window is selected for detection. b) photons statistics of the emitter in the PCS experiment: only red photons are detected c) HLAF in the case of an infinitely narrow linewidth}\label{infiniteline}
\end{figure}

We introduce now a detection energy condition such as we only detect photons emitted from the high energy side of the spectrum called right side (R side) in fig. \ref{infiniteline}. Therefore, we can discriminate the photons emitted from the high energy side of the spectrum (ie. R side, red spikes in fig. \ref{infiniteline} b) ) from the photons emitted in the low energy side (L side, blue spikes in fig. \ref{infiniteline} b) ). In the fig. \ref{infiniteline} b), bunches of single photons (red spikes) emitted from the R side are observed because the spectral diffusion is a time-correlated process (ie. with memory). On the HLAF (fig. \ref{infiniteline} c)), it results in a bunching combined with a zero delay dip due to the subpoissonian nature of the emission. If we consider that the emitter homogeneous linewidth is infinitely small, the HLAF can be easily calculated with a rate equation model by considering a split two-levels system with a probability for the emitter energy to switch from one side to the other one.

\begin{equation}
HLAF(\tau)=g^{2}(R,R,\tau)=[1+exp(-\gamma_{c}\tau)][1-exp(-(r+\gamma)\tau)]
\end{equation}

with $\gamma_{c} = \frac{1}{\tau_c}$, the correlation rate, r the pump rate and $\gamma$  the spontaneous emission rate.

It is the product of two expressions. Since $[1-exp(-(r+\gamma)\tau)]$
describes the single photon behavior of the emitter, $[1+exp(-\gamma_{c}\tau)]$
describes the spectral diffusion phenomenon and its influence on the
ability of the emitter to send photons in the right spectral window.
The spectral diffusion term is equal to 2 for the null delay.

\begin{figure}[h]
\noindent \centering{}\includegraphics[scale=0.4]{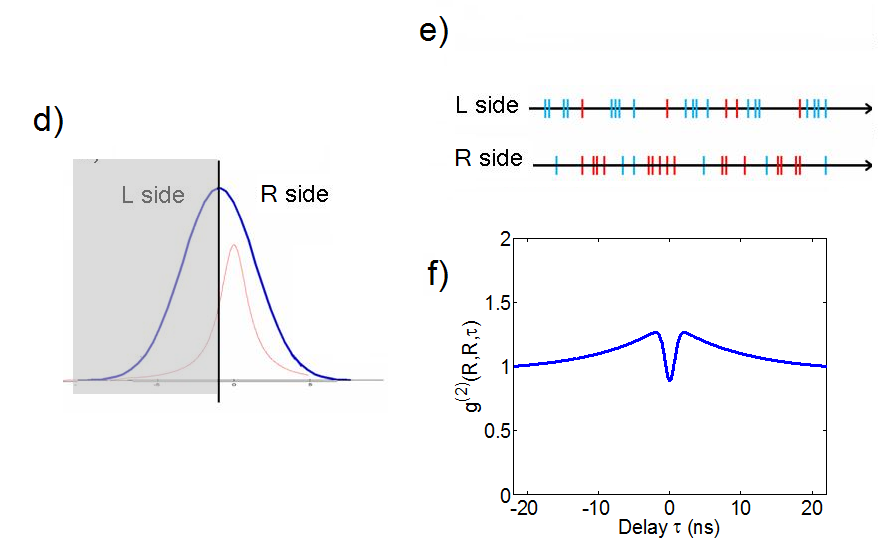}\caption{a) spectrum of a spectrally diffusing emitter with a finite linewidth: only the right spectral window is selected for detection. b) photons statistics of the emitter in the PCS experiment: only red photons are detected c) HLAF in the case of a non-negligible homogeneous linewidth}\label{finiteline}
\end{figure}

We now consider a finite homogeneous linewidth (fig. \ref{finiteline}) . 
The mean value of the homogeneous lineshape transits from one side of the spectral window to the other with the time constant $\tau_c$.
However, even if it is centered in the left side, a photon can be emitted in the right side and vice versa. The consequences on 
the photons statistics are represented in fig. \ref{finiteline} b). To the bunching statistics due to the time correlated spectral diffusion process
it has to be added an uncorrelated random distribution of the photons on both spectral windows, degrading the photons "bunching". The consequence on the HLAF is a diminution of the bunching contrast, as the energies of photons are less time correlated (fig. \ref{finiteline} c)).
The importance of this contrast is linked to the relative importance of the homogeneous linewidth compared
to the spectral diffusion amplitude. To understand this dependence, we calculate in the following the HLAF corresponding
to a spectrally diffusing single photon emitter with a finite homogeneous linewidth.

\section{Energy and emission separation}

The aim of this section is the obtention of a preliminary result,
which will simplify the  calculations leading to the
derivation of the HLAF for a non-negligible homogeneous linewidth.

We propose to show  that the HLAF is the product of the spectral diffusion part (carrying information
on the correlation of the emitter energy) and the single
photon part (carrying information on the correlation of the emission).
This  result only holds  when emission and
energy position of the line are independent. The emitter is seen as a 
two-level system (TLS) of fluctuating energy.

Let us define the following ensembles:

-$\gamma^{t}:$ ensemble of events such as a photon is detected at
time t.

- $\gamma_{em}^{t}$: ensemble of events such as a photon is emitted
by the TLS.

- $\gamma_{in}^{t}$: ensemble of events such as the TLS
energy is in the right spectral window.

The emission of a photon by the two level system (TLS) and its energy situated in the
right spectral window are two independent events, so:

\begin{equation}
\rho(\gamma_{em}^{\tau}\cap\gamma_{in}^{\tau})=\rho(\gamma_{em}^{\tau})\rho(\gamma_{in}^{\tau})\label{eq:26-1}
\end{equation}

We go back to the expression of autocorrelation function in term of
probabilities:

\begin{equation}
g^{2}(\tau)=\frac{\rho(\gamma^\tau\mid\gamma^0)}{\rho(\gamma^\tau)}
\end{equation}

To detect a photon at time t, one needs to have a photon emitted by
the TLS at time t and to have the TLS energy in the right spectral
window at time t:

\begin{equation}
\gamma^t=\gamma_{em}^{t}\cap\gamma_{in}^{t}
\end{equation}

Therefore,

\begin{equation}
\rho(\gamma^\tau\mid\gamma^0)=\rho(\gamma_{em}^{\tau}\cap\gamma_{in}^{\tau}\mid\gamma_{em}^{0}\cap\gamma_{in}^{0})\label{eq:etoui-1}
\end{equation}

Using (\ref{eq:26-1})  in (\ref{eq:etoui-1}), we have:

\begin{multline}
\rho(\gamma(\tau)\mid\gamma(0))=\rho(\gamma_{em}^{\tau}\mid\gamma_{em}^{0}\cap\gamma_{in}^{0})\rho(\gamma_{in}^{\tau}\mid\gamma_{em}^{0}\cap\gamma_{in}^{0})\\
=\rho(\gamma_{em}^{\tau}\mid\gamma_{em}^{0})\rho(\gamma_{in}^{\tau}\mid\gamma_{in}^{0})
\end{multline}

We can now find the general expression, separating the TLS
emission and energy correlation functions in the HLAF.

\begin{equation}
g^{2}(\tau)=\frac{\rho(\gamma_{em}^{\tau}\mid\gamma_{em}^{0})}{\rho(\gamma_{em}^{\tau})}\frac{\rho(\gamma_{in}^{\tau}\mid\gamma_{in}^{0})}{\rho(\gamma_{in}^{\tau})}=g_{em}^{2}(\tau)g_{in}^{2}(\tau) \label{eq:8Aurelien}
\end{equation}

This result will make our future calculations of HLAF
easier, since the emission part $g_{em}^{2}(\tau)$
will always be the same and we will only have to focus on the derivation
of $g_{in}^{2}(\tau)$, the second order correlation function of the
TLS energy position.

\section{Half-line autocorrelation function}

Since the mean TLS energy fluctuates from the right
side to the other with a correlated statistics, but can also be randomly,
and with no time-correlation, distributed in energy along the lorentzian
profile imposed by the homogeneous linewidth, The total energy $E^{t}$of the TLS at time t is the sum of two random
variables:

\begin{equation}
E^{t}=\mu^{t}+\varepsilon^{t}
\end{equation}

where $\mu^{t}$ is the energy position of the center of the homogeneous
lineshape at time t. It is distributed along the gaussian distribution
of the fluctuations \cite{kubo, anderson}. This random variable is time correlated. It means that
the value of the random variable at time $t+\tau$ is influenced by
its value at time t and :
\begin{equation}
\langle\mu^{t}\mu^{t+\tau}\rangle\neq\langle\mu^{t}\rangle\langle\mu^{t+\tau}\rangle
\end{equation}

$\varepsilon^{t}$ is the energy shift due to the homogeneous linewidth
at time t. We assume it has a lorentzian distribution centered on
$\mu^{t}$. This variable is described by a poissonian process and
is not time-correlated. 
\begin{equation}
\langle\varepsilon^{t}\varepsilon^{t+\tau}\rangle=\langle\varepsilon^{t}\rangle\langle\varepsilon^{t+\tau}\rangle=\langle\varepsilon^{0}\rangle\langle\varepsilon^{\tau}\rangle=\langle\varepsilon^{\tau}\rangle^{2}
\end{equation}

The spectral window of detection is defined by the energy interval
$I_{n}$. In this letter, we consider, $I_{n}=[0,+\infty[$, ie. corresponding to the R side.

The energy of the TLS is in the right spectral window at time t when
$E^{t}\epsilon I_{n}$, so when:
\begin{equation}
(\varepsilon^{t}+\mu^{t})\epsilon I_{n}
\end{equation}

Therefore, we have:
\begin{align*}
\gamma_{in}^{t}\equiv\left\{ (\varepsilon^{t}+\mu^{t})\epsilon(I_{n})\right\} \\
\gamma_{in}^{0}\equiv\left\{ (\varepsilon^{0}+\mu^{0})\epsilon(I_{n})\right\} \\
\end{align*}

We also define the following ensembles : 

-$\mu_{in}^{t}$ : ensemble of $\mu$ such as $\mu^{t}\epsilon I_{n}$. 

-$\mu_{out}^{t}$ : ensemble of $\mu$ such as $\mu^{t}\notin I_{n}$ 

We found out in the previous subsection that we can always write down
the autocorrelation function as:
\[
g^{2}(\tau)=g_{em}^{2}(\tau)g_{in}^{2}(\tau)
\]

with $g_{in}^{2}(\tau)=\frac{\rho(\gamma_{in}^{\tau}\mid\gamma_{in}^{0})}{\rho(\gamma_{in}^{\tau})^2}$.

\begin{equation}
\rho(\gamma_{in}^{\tau}\mid\gamma_{in}^{0})=\rho(\left\{ (\varepsilon^{\tau}+\mu^{\tau})\epsilon(I_{n})\right\} \mid\gamma_{in}^{0})\label{eq:27}
\end{equation}

We use here the law of total probability which asserts that
for an ensemble A and its partition $\left\{A_i\right\}$ such as $A=\sum_{i}A_{i}$,

\[
P(B)=\sum_{i}P(B\mid A_{i})P(A_{i})
\]

So, eq. (\ref{eq:27}) becomes:
\begin{multline}
\rho(\gamma_{in}^{\tau}\mid\gamma_{in}^{0})=\rho(\left\{ (\varepsilon^{\tau}+\mu^{\tau})\epsilon(I_{n})\right\} \mid\mu_{in}^{\tau}\mid\gamma_{in}^{0})\rho(\mu_{in}^{\tau}\mid\gamma_{in}^{0})\\
+\rho(\left\{ (\varepsilon^{\tau}+\mu^{\tau})\epsilon(I_{n})\right\} \mid\mu_{out}^{\tau}\mid\gamma_{in}^{0})\rho(\mu_{out}^{\tau}\mid\gamma_{in}^{0})\\
\end{multline}

$(\left\{ (\varepsilon^{\tau}+\mu^{\tau})\epsilon(I_{n})\right\} \mid\mu_{in}^{\tau})$,
the ensemble of values taken by $\varepsilon$ and $\mu$ such as
$(\varepsilon+\mu)\epsilon(I_{n})$ at time $\tau$, knowing that
$\mu\epsilon I_{n}$ at time $\tau$ is independent from the possible
values taken by $\varepsilon$ at time 0. Indeed, $\varepsilon$ is
a poissonnian random process. Moreover, the condition on the $\mu$ value is
already fixed at time $\tau$ in $(\left\{ (\varepsilon^{\tau}+\mu^{\tau})\epsilon(I_{n})\right\} \mid\mu_{in}^{\tau})$,
a condition on its value at time 0 does not change the ensemble.

So we can conclude that the ensemble $(\left\{ (\varepsilon^{\tau}+\mu^{\tau})\epsilon(I_{n})\right\} \mid\mu_{in}^{\tau})$
and $\gamma_{in}^{0}$ are independent ensembles, thus:
\begin{multline}
\rho(\left\{ (\varepsilon^{\tau}+\mu^{\tau})\epsilon(I_{n})\right\} \mid\mu_{in}^{\tau}\mid\gamma_{in}^{0})\rho(\mu_{in}^{\tau}\mid\gamma_{in}^{0})\\
=\rho(\left\{ (\varepsilon^{\tau}+\mu^{\tau})\epsilon(I_{n})\right\} \mid\mu_{in}^{\tau})\rho(\mu_{in}^{\tau}\mid\gamma_{in}^{0})\label{eq:28}
\end{multline}

and 
\begin{multline}
\rho(\gamma_{in}^{\tau}\mid\gamma_{in}^{0})=\rho(\left\{ (\varepsilon^{\tau}+\mu^{\tau})\epsilon(I_{n})\right\} \mid\mu_{in}^{\tau})\rho(\mu_{in}^{\tau}\mid\gamma_{in}^{0})\\
+\rho(\left\{ (\varepsilon^{\tau}+\mu^{\tau})\epsilon(I_{n})\right\} \mid\mu_{out}^{\tau})\rho(\mu_{out}^{\tau}\mid\gamma_{in}^{0})\label{eq:16Aurelien}
\end{multline}

For any ensembles A and B, one have the basic relationship:
\[
P(A\mid B)=P(B\mid A)\frac{P(A)}{P(B)}
\]

so (\ref{eq:16Aurelien}) becomes:
\begin{multline}
\rho(\gamma_{in}^{\tau}\mid\gamma_{in}^{0})=\rho(\left\{ (\varepsilon^{\tau}+\mu^{\tau})\epsilon(I_{n})\right\} \mid\mu_{in}^{\tau})\rho(\gamma_{in}^{0}\mid\mu_{in}^{\tau})\frac{\rho(\mu_{in}^{\tau})}{\rho(\gamma_{in}^{0})}\\
+\rho(\left\{ (\varepsilon^{\tau}+\mu^{\tau})\epsilon(I_{n})\right\} \mid\mu_{out}^{\tau})\rho(\gamma_{in}^{0}\mid\mu_{out}^{\tau})\frac{\rho(\mu_{out}^{\tau})}{\rho(\gamma_{in}^{0})}\label{eq:29}
\end{multline}

We apply again the law of total probability on the terms $\rho(\gamma_{in}^{0}\mid\mu_{in}^{\tau})$
and $\rho(\gamma_{in}^{0}\mid\mu_{out}^{\tau})$.

\begin{multline}
\rho(\gamma_{in}^{0}\mid\mu_{in}^{\tau})=\rho((\gamma_{in}^{0}\mid\mu_{in}^{0})\mid\mu_{in}^{\tau})\rho(\mu_{in}^{0}\mid\mu_{in}^{\tau})\\
+\rho((\gamma_{in}^{0}\mid\mu_{out}^{0})\mid\mu_{in}^{\tau})\rho(\mu_{out}^{0}\mid\mu_{in}^{\tau})\\
=\rho(\gamma_{in}^{0}\mid\mu_{in}^{0})\rho(\mu_{in}^{0}\mid\mu_{in}^{\tau})\\
+\rho(\gamma_{in}^{0}\mid\mu_{out}^{0})\rho(\mu_{out}^{0}\mid\mu_{in}^{\tau})\\
=\rho(\gamma_{in}^{0}\mid\mu_{in}^{0})\rho(\mu_{in}^{\tau}\mid\mu_{in}^{0})\frac{\rho(\mu_{in}^{0})}{\rho(\mu_{in}^{\tau})}\\
+\rho(\gamma_{in}^{0}\mid\mu_{out}^{0})\rho(\mu_{in}^{\tau}\mid\mu_{out}^{0})\frac{\rho(\mu_{out}^{0})}{\rho(\mu_{in}^{\tau})}
\end{multline}

The same way, 
\begin{multline}
\rho(\gamma_{in}^{0}\mid\mu_{out}^{\tau})=\rho(\gamma_{in}^{0}\mid\mu_{in}^{0})\rho(\mu_{out}^{\tau}\mid\mu_{in}^{0})\frac{\rho(\mu_{in}^{0})}{\rho(\mu_{out}^{\tau})}\\
+\rho(\gamma_{in}^{0}\mid\mu_{out}^{0})\rho(\mu_{out}^{\tau}\mid\mu_{out}^{0})\frac{\rho(\mu_{out}^{0})}{\rho(\mu_{out}^{\tau})}
\end{multline}

We inject the two last equations in (\ref{eq:29}):
\begin{multline}
\rho(\gamma_{in}^{\tau}\mid\gamma_{in}^{0})=\rho(\gamma_{in}^{\tau}\mid\mu_{in}^{\tau})\rho(\gamma_{in}^{0}\mid\mu_{in}^{0})\rho(\mu_{in}^{\tau}\mid\mu_{in}^{0})\frac{\rho(\mu_{in}^{0})}{\rho(\gamma_{in}^{0})}\\
+\rho(\gamma_{in}^{\tau}\mid\mu_{in}^{\tau})\rho(\gamma_{in}^{0}\mid\mu_{out}^{0})\rho(\mu_{in}^{\tau}\mid\mu_{out}^{0})\frac{\rho(\mu_{out}^{0})}{\rho(\gamma_{in}^{0})}\\
+\rho(\gamma_{in}^{\tau}\mid\mu_{out}^{\tau})\rho(\gamma_{in}^{0}\mid\mu_{in}^{0})\rho(\mu_{out}^{\tau}\mid\mu_{in}^{0})\frac{\rho(\mu_{in}^{0})}{\rho(\gamma_{in}^{0})}\\
+\rho(\gamma_{in}^{\tau}\mid\mu_{out}^{\tau})\rho(\gamma_{in}^{0}\mid\mu_{out}^{0})\rho(\mu_{out}^{\tau}\mid\mu_{out}^{0})\frac{\rho(\mu_{out}^{0})}{\rho(\gamma_{in}^{0})}\\
\label{eq:30}
\end{multline}

$\rho(\mu_{in}^{0})$, $\rho(\mu_{out}^{0})$and
$\rho(\gamma_{in}^{0})$ are respectively the probability to have
$\mu\in I_{n},$ $\mu\notin I_{n}$, and $\left\{ (\varepsilon^{\tau}+\mu^{\tau})\epsilon(I_{n})\right\} $
at null time. Thus:
\[
\rho(\mu_{in}^{0})=\rho(\mu_{out}^{0})=\rho(\gamma_{in}^{0})=\frac{1}{2}
\]

and (\ref{eq:30}) becomes
\begin{multline}
\rho(\gamma_{in}^{\tau}\mid\gamma_{in}^{0})=\rho(\gamma_{in}^{\tau}\mid\mu_{in}^{\tau})\rho(\gamma_{in}^{0}\mid\mu_{in}^{0})\rho(\mu_{in}^{\tau}\mid\mu_{in}^{0})\\
+\rho(\gamma_{in}^{\tau}\mid\mu_{in}^{\tau})\rho(\gamma_{in}^{0}\mid\mu_{out}^{0})\rho(\mu_{in}^{\tau}\mid\mu_{out}^{0})\\
+\rho(\gamma_{in}^{\tau}\mid\mu_{out}^{\tau})\rho(\gamma_{in}^{0}\mid\mu_{in}^{0})\rho(\mu_{out}^{\tau}\mid\mu_{in}^{0})\\
+\rho(\gamma_{in}^{\tau}\mid\mu_{out}^{\tau})\rho(\gamma_{in}^{0}\mid\mu_{out}^{0})\rho(\mu_{out}^{\tau}\mid\mu_{out}^{0})\\
\label{eq:31-1}
\end{multline}

The probability for the TLS to be in a the right spectral window,
knowing that it was in it at t=0, is the sum of four probabilities, describing
the four possible configurations for the occurrence of a coincidence:

-For the first term, the homogeneous line is in the R spectral window
at t=0 and also at t=$\tau$.

-For the second, the homogeneous line is out of the R spectral window
at t=0 and in at t=$\tau$

-For the third, the homogeneous line is in the R spectral window at
t=0 and out at t=$\tau$.

-For the fourth term, the homogeneous line is out of the R spectral
window at t=0 and also at t=$\tau$.

Let us evaluate $\rho(\gamma_{in}^{\tau}\mid \mu_{in(out)}^{\tau})$,
 the probability that the TLS energy is in the right spectral window
at time $\tau$ when the center of the homogeneous line is in (out of) the
right spectral window. Despite the fact that the homogeneous linewidth is of relative complex shape \cite{lucien}, we approximate it as  a lorentzian centered on $E=\mu$.
In accordance with the Kubo-Anderson model, the energy distribution
of the fluctuations is gaussian \cite{kubo, anderson}. We have:

\begin{equation}
\rho(\gamma_{in}^{\tau}\mid\mu_{in(out)}^{\tau})=\frac{\rho(\gamma_{in}^{\tau}\cap\mu_{in(out)}^{\tau})}{\rho(\mu_{in(out)}^{\tau})}=2\rho(\gamma_{in}^{\tau}\cap\mu_{in(out)}^{\tau})
\end{equation}

We consider a Lorentzian centered in the right spectral window (ie. $\mu\epsilon I_{in}$),
and calculate the probability to measure a photon emitted in
the right spectral window.

\begin{figure}[h]
\noindent \begin{centering}
\includegraphics[scale=0.35]{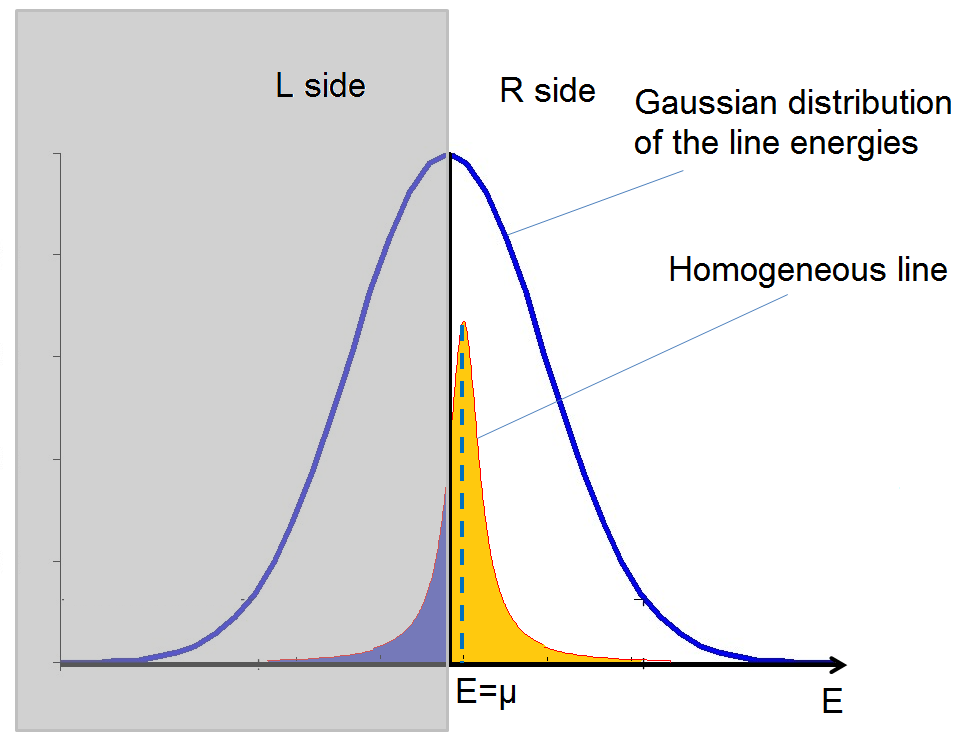}\label{lorentzgauss}
\par\end{centering}

\caption{Lorentzian centered on energy $\mu$ and gaussian
distribution of the possible $\mu$ positions. This
distribution is cut by the spectral selection of the right part of
the inhomogeneous line. It defines the area of the lorentzian where
photons can be detected.}

\end{figure}

For a given lorentzian centered in $\mu$, the probability for the
TLS to emit a photon in the right spectral window is the ratio between the lorentzian
area in the right spectral window and its total area:

\begin{equation}
P_{\mu}(\gamma_{in}^{\tau})=\frac{1}{\int^{\infty}lor(\sigma,E-\mu) dE}\int_{E\epsilon I_{in}}lor(\sigma,E-\mu) dE
\end{equation}

with 
\[
lor(\sigma,E-\mu)=\frac{\frac{2}{\pi\sigma}}{1+(\frac{E-\mu}{\sigma/2})^{2}}
\]

We have then to consider the probability for the lorentzian to be
centered at the energy E between $\mu$ and $\mu + d\mu$. It is obtained in multiplying the previous
expression by the probability density imposed by the gaussian distribution
$Gauss(\Sigma,\mu)d\mu$ and integrating over all $\mu$ such
as $\mu\epsilon I_{in}$.

\begin{multline}
\rho(\gamma_{in}^{\tau}\cap\mu_{in(out)}^{\tau})=\\
\frac{\int_{\mu\epsilon I_{in(out)}} Gauss(\Sigma,\mu)\int_{E\epsilon I_{in}}(lor(\sigma,E-\mu) dE) d\mu}{\int^{\infty}_0 Gauss(\Sigma,\mu)\int^{\infty}_0(lor(\sigma,E-\mu) dE) d\mu}
\end{multline}

with $Gauss(\Sigma,\mu)=\frac{1}{\Sigma_{s} \sqrt{2\pi}}exp\left(\frac{-\mu^{2}}{2\Sigma_{s}^{2}}\right)$, $\Sigma_{s}=\frac{\Sigma}{2\sqrt{2ln2}}$ the standard deviation and $\Sigma$ the full width at half maximum of the gaussian function.

This is the probability for the TLS to be in the right spectral window
and the lorentzian centered in the right (left) spectral window at
any time, so it does not depend on the delay $\tau$ and are only functions of the
homogeneous linewidth $\sigma$ and the gaussian standard
deviation $\Sigma_S$: 
\begin{equation}
\rho(\gamma_{in}^{\tau}\cap\mu_{in(out)}^{\tau})=\rho(\gamma_{in}^{0}\cap\mu_{in(out)}^{0})=\alpha_{in(out)}(\sigma,\Sigma)
\end{equation}

We rewrite the eq. \ref{eq:31-1} with these coefficients:

\begin{multline}
\rho(\gamma_{in}^\tau \mid \gamma_{in}^0) = 4 [\alpha_{in}^{2}\rho(\mu_{in}^{\tau}\mid\mu_{in}^{0})\\
+\alpha_{in}\alpha_{out}\rho(\mu_{in}^{\tau}\mid\mu_{out}^{0})\\
+\alpha_{out}\alpha_{in}\rho(\mu_{out}^{\tau}\mid\mu_{in}^{0})\\
+\alpha_{out}^{2}\rho(\mu_{out}^{\tau}\mid\mu_{out}^{0})]\\
\label{eq:32}
\end{multline}

This probability depends on one hand on the coefficients $\alpha_{in}$ and
$\alpha_{out}$ which depend themselves on the linewidth of the homogeneous
lorentzian line and of the gaussian distribution. They take into
account the random uncorrelated statistics introduced by the finite
homogeneous linewidth. On the other hand, eq. (\ref{eq:32}) depends on
the probabilities $\rho(\mu_{in}^{\tau}\mid\mu_{in}^{0})$, $\rho(\mu_{in}^{\tau}\mid\mu_{out}^{0}),$$\rho(\mu_{out}^{\tau}\mid\mu_{in}^{0})$,
$\rho(\mu_{out}^{\tau}\mid\mu_{out}^{0})$ which are describing how
the mean value of the homogeneous linewidth is transiting from one spectral
window to the other and take into account the time-correlated part of the
statistics. All these probabilities can be determined analytically
with the model of the infinitely sharp homogeneous linewidth presented
in the reference \cite{gregdiff}, and its associated rate equations.

Because of the configuration chosen in this letter (ie. left side versus right side of the spectrum), the probability $\gamma_R$ for the mean value of the homogeneous line to jump from the right spectral window to the left is the same as the probability $\gamma_L$ for the homogeneous line 
to make the opposite move, thus:

\begin{equation}
 \gamma_{R}=\gamma_{L}=\frac{\gamma_{c}}{2}
\end{equation}

Then, we can derive:
\begin{equation}
\rho(\mu_{in}^{\tau}\mid\mu_{in}^{0})=\frac{\gamma_{R}}{\gamma_{c}}+(1-\frac{\gamma_{R}}{\gamma_{c}})exp(-\gamma_{c}\tau)=\frac{1}{2} \left( 1 + exp(-\gamma_{c}\tau)\right)
\end{equation}

\begin{multline}
\rho(\mu_{in}^{\tau}\mid\mu_{out}^{0}) = \frac{\gamma_{R}}{\gamma_{c}}[1-exp(-\gamma_{c}\tau)]\\
=\frac{1}{2}[1-exp(-\gamma_{c}\tau)]\\
\rho(\mu_{out}^{\tau}\mid\mu_{in}^{0})=\frac{1}{2}[1-exp(-\gamma_{c}\tau)]\\
\rho(\mu_{out}^{\tau}\mid\mu_{out}^{0})=\frac{1}{2}[1+exp(-\gamma_{c}\tau)]\\
%\\=[\frac{\gamma_{L}}{\gamma_{c}}+(1-\frac{\gamma_{L}}{\gamma_{c}})exp(-\gamma_{c}\tau)]\\
\label{eq:grege}
\end{multline}

We put all these expressions in eq. (\ref{eq:32}) :

\begin{multline}
\rho(\gamma_{in}^\tau \mid\gamma_{in}^0)=4[\frac{1}{2}(\alpha_{in}^{2}+\alpha_{out}^{2})+\alpha_{in}\alpha_{out}\\
+(\frac{1}{2}(\alpha_{in}^{2}+\alpha_{out}^{2})-\alpha_{in}\alpha_{out}).exp(-\gamma_{c}\tau)]
\end{multline}

To obtain the correlation function, we divide this expression by $\rho(\gamma_{in}^\infty\mid\gamma_{in}^0)=4[\frac{1}{2}(\alpha_{in}^{2}+\alpha_{out}^{2})+\alpha_{in}\alpha_{out}]$,

\begin{equation}
g_{in}^{2}(\tau)=1+\frac{(\frac{1}{2}(\alpha_{in}^{2}+\alpha_{out}^{2})-\alpha_{in}\alpha_{out})}{(\frac{1}{2}(\alpha_{in}^{2}+\alpha_{out}^{2})+\alpha_{in}\alpha_{out})} exp(-\gamma_{c}\tau)
\end{equation}

finally, rearranging terms and multiplying the emission part $g_{em}^{2}(\tau)$,
one has:

\begin{multline}
g^{2}(R,R,\tau)=[1+\frac{(\alpha_{in}-\alpha_{out})^{2}}{(\alpha_{in}+\alpha_{out})^{2}}.exp(-\gamma_{c}\tau)]\\
*[1-exp(-(r+\gamma)\tau)]\label{eq:analty}
\end{multline}

 Thus, the HLAF is on the form:
\begin{equation}
g^{2}(R,R,\tau)=[1+\beta.exp(-\gamma_{c}\tau)][1-exp(-(r+\gamma)\tau)]\label{eq:analty}
\end{equation}

with,
\begin{equation}
\beta=\frac{(\alpha_{in}-\alpha_{out})^{2}}{(\alpha_{in}+\alpha_{out})^{2}}\label{eq:35Aurelien}
\end{equation}

\begin{figure}[h]

\includegraphics[scale=0.3]{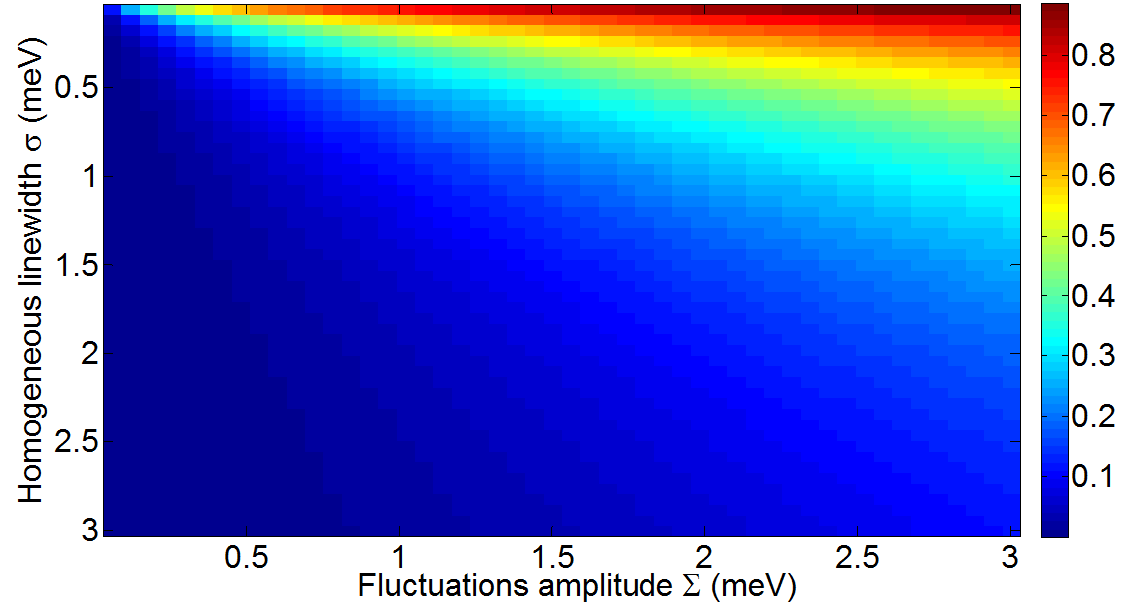}\caption{Bunching factor $\beta$ as a function of the fluctuation amplitude $\Sigma$, and the homogeneous
linewidth $\sigma$}\label{bunchdependance}
\end{figure}

We can make two remarks about this factor:

- The bunching factor $\beta$ is exclusively ruled by the uncorrelated
random statistics of the finite homogeneous linewidth (expressed by
coefficients $\alpha_{in}$and $\alpha_{out}$), and is a function
of $\sigma$ and $\Sigma$. As shown in fig. \ref{bunchdependance}, the
bunching is important (red color) for small $\sigma$ and large fluctuation amplitudes $\Sigma$. 

- The behavior of $\beta$ at the limits is interesting: when the
homogeneous linewidth tends to 0, ie. $\sigma\ll\Sigma$, we have:
\begin{align*}
\alpha_{in}\rightarrow\frac{1}{2}\\
\alpha_{out}\rightarrow0
\end{align*}

so, $\beta\rightarrow1$, which is its expected value in the case
of the infinitely sharp linewidth.

When the homogeneous linewidth becomes  larger than the standard
deviation of the fluctuation, $\sigma\gg\Sigma$:
\begin{align*}
\alpha_{in}\rightarrow\frac{1}{4}\\
\alpha_{out}\rightarrow\frac{1}{4}\\
\end{align*}

and $\beta\rightarrow0$, the bunching is collapsing. This situation corresponds
to a case in which the poissonian statistics of the homogeneous linewidth
takes over the correlated statistics of the spectral diffusion.

 The HLAF bears simultaneously the signature of the subpoissonian emission
statistics (zero delay dip), the correlated spectral diffusion energy
statistics (bunching), and the poissonian energy statistics of the homogeneous
linewidth (degradation of the bunching), the two last signatures having opposing effects.

\section{Monte-Carlo simulation}

In order to confirm the analytical expression (\ref{eq:analty})
derived from the calculations above, we performed a Montecarlo simulation,
by building numerically the function $g^{2}(R,R,\tau)$ for an emitter spectrally diffusing in and out of the detection spectral
window. We do not take into account the subpoissonian statistics of the emitted since it can be factorized in the correlation function (see eq. \ref{eq:8Aurelien}). Thus, the simulation gives a direct access to $g_{in}^{2}(\tau)$
and consequently to $\beta$ since

\begin{equation}
\beta=g_{in}^{2}(0)
\end{equation}

After generating a poissonian stream of photons, we assign to each of
them a random energy in a gaussian distribution. We calculate the
probability of the mean value of the homogeneous line to jump between the emission for
the $(i-1)^{th}$ photon and the $i^{th}$ photon.

The energy of the $(i-1)^{th}$ photon is in the interval:

\[
I_{i-1}=I_{in}=\left[E_{i-1}-\frac{\delta E}{2},E_{i-1}+\frac{\delta E}{2}\right]
\]

with $\delta E$ an infinitesimal energy.

In the case of a jump of the homogeneous line, the energy of the $i^{th}$
photon is in the interval:

\[
I_{i}=I_{out}=\left[-\infty,E_{i-1}-\frac{\delta E}{2}\right]\cup\left[E_{i-1}+\frac{\delta E}{2},+\infty\right]
\]

Thus, the probability for the homogeneous line to jump between the
emission of the ith and the $(i-1)^{th}$ photon is:

\begin{multline}
P_{jump(i)}=\rho(\left\{ \mu^{\tau}\epsilon I_{out}\right\} \mid\left\{ \mu^{0}\epsilon I_{in}\right\} )=\frac{\gamma_{out}}{\gamma_{c}}[1-exp(-\gamma_{c}\tau)]
\end{multline}

This probability has already been calculated in the previous section (see equation (\ref{eq:grege})). 

$\gamma_{out}$ is the exit rate of the homogeneous line, from $I_{in}$ to $I_{out}$,$\gamma_{c}$ is the {}``jump rate'' of the homogeneous
line.

For an infinitely small $\delta E$, $\gamma_{c}\backsim\gamma_{out}$
and:
\[
P_{jump(i/(i-1))}=[1-exp(-\gamma_{c}\tau)]
\]

If the $i^{th}$ photon does not jump, it takes the same energy than the
$(i-1)^{th}$ photon.

We then discriminate the photons belonging to energies outside the detection
area by applying an energy condition. They will not be counted in the
correlation process. 

To calculate the correlation function of the resulting photons stream,
we compute the delay between the arrival of each photon and all the
other photons of the stream. By building the histogram of these delays,
one obtains the correlation function of the stream \cite{streamcorr}.

\begin{figure}[h]
\noindent \centering{}\includegraphics[scale=0.25]{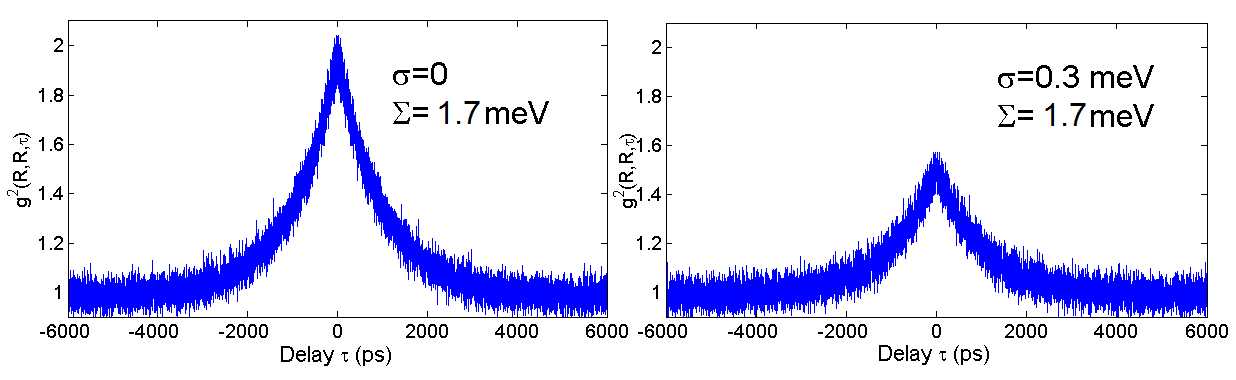}\caption{Calculated half line autocorrelation functions of a poissonnian emitter spectrally diffusing.
a) for an infinitely small homogeneous linewidth, b) for a homogeneous
linewidth $\sigma=0.3meV$.}
\label{bunchsimu}
\end{figure}

The result of the simulation is plotted on fig. \ref{bunchsimu} a). We
find again the result explained in the previous section, the bunching
part of the HLAF takes a value of 2
at zero delay. To evaluate $\beta$ from the simulated data, one only
need to extract $g^{2}(0)$.

To simulate the finite homogeneous linewidth effect, we add for each
photon an energy shift $\varepsilon$. $\varepsilon$ is a random
variable distributed along a lorentzian distribution of linewidth
$\sigma$. The effect of this addition is shown on fig. \ref{bunchsimu} b).
As expected the bunching is less marked and the $\beta$ factor
goes from 1 to 0.5 in the case of a finite homogeneous linewith of $\sigma=0.3meV$
and a fluctuation amplitude of $\Sigma=1.7\, meV$.

\begin{figure}[h]
\noindent \centering{}\includegraphics[scale=0.3]{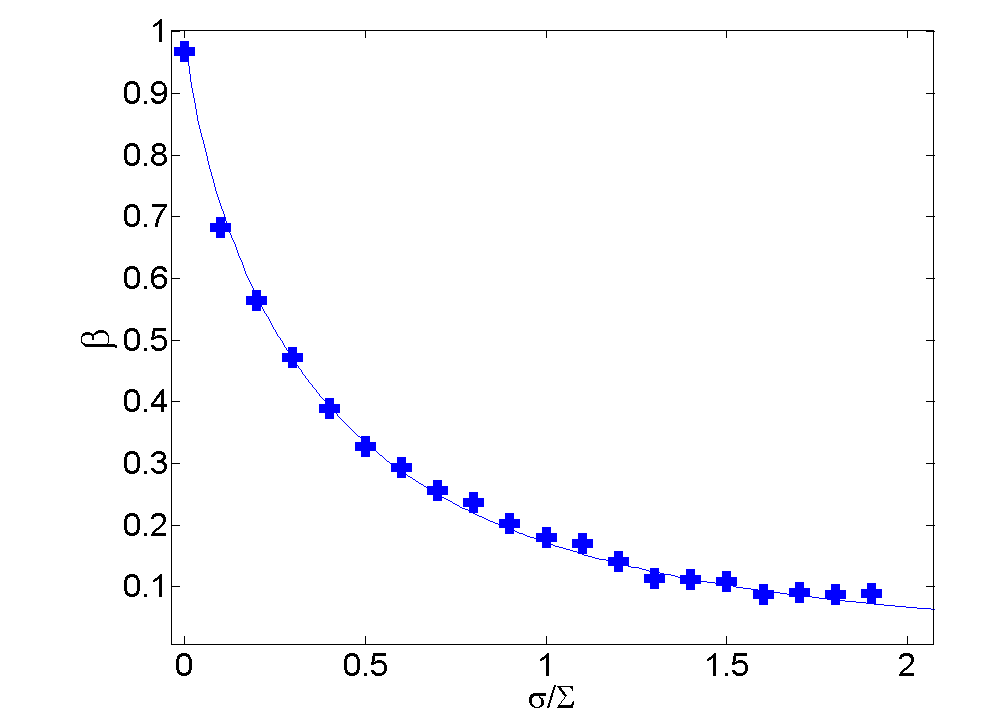}\caption{$\beta$ factor versus the ratio between homogeneous linewidth $\sigma$ and fluctuation
amplitude $\Sigma$ for fixed value of $\Sigma$. Red line: analytical model described by eq. \ref{eq:35Aurelien}. Blue
dots: Monte Carlo simulation result.}\label{monteanalyt}
\end{figure}

To compare this simulation with the analytical expression, we fix
the $\Sigma$ parameter and change $\sigma$. We then report the values of the $\beta$ factor versus the ratio $\frac{\sigma}{\Sigma}$ in both cases (see fig. \ref{monteanalyt}).
We can notice that the two methods (analytical and Monte Carlo) give
the same dependence.

\section{Homogeneous and inhomogeneous linewidth determination}

The fit of the experimental data allows to extract, as two independent parameters, the correlation time $\tau_{c}$ and $\beta$. As
shown in the previous section the latter depends on the fluctuation
amplitude $\Sigma$ and the homogeneous linewidth $\sigma$. To estimate
them separately, we can use the emission spectrum measured, and by
making the same assumptions as before (the homogeneous line is lorentzian
and the fluctuation distribution is gaussian) we obtain a second equation
linking these two parameters since, in this case, the emission spectrum has a Voigt profile:

\begin{equation}
S(\sigma,\Sigma,\omega)=\intop_{-\infty}^{+\infty}lor(\sigma,\xi).gauss(\Sigma,\xi-\omega)d\xi \label{eq:36}
\end{equation}

By a double fit, one can find the parameters couple $(\sigma,\Sigma)$
which is satisfying these two equations, giving access to the amplitude of the spectral diffusion and the homogeneous linewidth separately.

\section{Conclusion}

The HLAF is not only a function of the correlation time of the spectral diffusion. It is dependent on the nature of the
photon energy statistics. From the calculation we performed, one can interpret analytically the importance of these 
different statistics in the observed correlation function. By measuring the HLAF and by making a temperature dependence of a spectrally diffusing emitter it is possible to describe how electronic fluctuations and phonon broadening are evolving, and which one is dominating the spectrum.Thus, PCS technique can bring all the
informations for the full characterization of a spectrally diffusing
emitter. Indeed, after data treatment technique presented in the last
section of this letter, one can obtain separately $\sigma$ (homogeneous
linewidth), $\Sigma$ (fluctuation amplitude) and $\gamma_{c}$ (correlation
rate). 
Other existing methods \cite{holeburn, spectre} , except the the photon correlation
Fourier spectroscopy method (PCFS) \cite{hermier1, hermier2} are not bringing all these informations at the same time and all of them are not
adapted in the case of fast fluctuations.
From an experimentalist point of view, it is important to note that the spectral resolution of
the PCS experiment is limited by the spectrometer, which makes it less accurate than 
PCFS , which takes advantage of the Fourier transform spectroscopy.
However the PCS technique is extremely easy to setup and does not require the drastic stability conditions
demanded by the PCFS technique. Since photons correlation experiments require long integration time, such sensitivity led to the degradation of the time resolution to 20 $\mu s$ for the PCFS technique, which forbids to probe fast fluctuations of the energy emitter. In PCS technique, time resolution is only limited by photodiodes and can be lowered to around 100 ps. This technique is a reliable and accessible technique for a complete and fast characterization of a single photon emitter in a solid state environment.

\end{document}